\documentstyle[12pt]{article}

\font\blackboard=msbm10 at 12pt
\font\blackboards=msbm7
\font\blackboardss=msbm5
\newfam\black
\textfont\black=\blackboard
\scriptfont\black=\blackboards
\scriptscriptfont\black=\blackboardss

\newcommand{\junk}[1]{}

\newcommand{\ba}{\begin{array}}
\newcommand{\ea}{\end{array}}
\newcommand{\be}{\begin{equation}}
\newcommand{\ee}{\end{equation}}
\newcommand{\bea}{\begin{eqnarray}}
\newcommand{\eea}{\end{eqnarray}}
\newcommand{\beas}{\begin{eqnarray*}}
\newcommand{\eeas}{\end{eqnarray*}}

\def\laplace{{\kern1pt\vbox{\hrule height 1.2pt\hbox{\vrule width
1.2pt\hskip
  3pt\vbox{\vskip 6pt}\hskip 3pt\vrule width 0.6pt}\hrule height
  0.6pt}
  \kern1pt}}
\def\scriptlap{{\kern1pt\vbox{\hrule height 0.8pt\hbox{\vrule width
  0.8pt
  \hskip2pt\vbox{\vskip 4pt}\hskip 2pt\vrule width 0.4pt}\hrule height
  0.4pt}
  \kern1pt}}

\def\roughly#1{\raise.3ex\hbox{$#1$\kern-
.75em\lower1ex\hbox{$\sim$}}}

\def\str{{\rm STr} \,}

\newcommand{\NP}{{\em Nucl.\ Phys.\ }}

\newcommand{\PL}{{\em Phys.\ Lett.\ }}

\newcommand{\PRL}{{\em Phys.\ Rev.\ Lett.\ }}

\newcommand{\gone}[1]{}

\begin{document}
\pagestyle{plain}
\setcounter{page}{1}

\baselineskip16pt

\begin{titlepage}

\begin{flushright}
PUPT-1940\\
MIT-CTP-3004\\
hep-th/0007157
\end{flushright}
\vspace{8 mm}

\begin{center}

{\Large \bf D-particle polarizations with multipole moments of 
higher-dimensional branes\\}

\end{center}

\vspace{7 mm}

\begin{center}

Karl Millar$^a$, Washington Taylor$^a$ and Mark Van Raamsdonk$^b$

\vspace{3mm}
${}^a${\small \sl Center for Theoretical Physics} \\
{\small \sl MIT, Bldg.  6-306} \\
{\small \sl Cambridge, MA 02139, U.S.A.} \\
{\small \tt kmillar@mit.edu, wati@mit.edu}\\

\vspace{3mm}
${}^b${\small \sl Department of Physics} \\
{\small \sl Joseph Henry Laboratories} \\
{\small \sl Princeton University} \\
{\small \sl Princeton, New Jersey 08544, U.S.A.} \\
{\small \tt mav@princeton.edu}
\end{center}

\vspace{8 mm}

\begin{abstract}
We study the polarization states of the D0-brane in type IIA string
theory.  In addition to states with angular momentum and
magnetic dipole moments, there are polarization states of a single
D0-brane with nonzero D2-brane dipole and magnetic H-dipole moments, as 
well as quadrupole and higher moments of various charges. These 
fundamental moments of the D0-brane polarization states can be 
determined directly from the linearized couplings of
background fields to the D0-brane world-volume fermions. 
These couplings determine the long range supergravity fields produced by 
a general polarization state, which typically have non-zero values for 
all the bosonic fields of type IIA supergravity. We demonstrate 
the precise cancellation between spin-spin and magnetic
dipole-dipole interactions and an analogous cancellation between
3-form and H-field dipole-dipole interactions for a pair of D0-branes.  
The first of these cancellations follows from the fact that spinning 
D0-brane states have gyromagnetic ratio $g = 1$, and the second follows 
from the fact that the ratio between the 3-form and H-field dipole 
moments is also 1 in natural units.  Both of these relationships can be 
seen immediately from the couplings in the D0-brane world-volume action.
\end{abstract}

\vspace{0.7cm}
\begin{flushleft}
July 2000
\end{flushleft}
\end{titlepage}
\newpage

\section{Introduction}

Understanding the relationship between black brane solutions of
supergravity carrying R-R charges and Dirichlet branes in string
theory has been one of the principal themes in the second superstring
revolution, following Polchinski's seminal paper relating these
objects \cite{Polchinski}.  A substantial literature is now devoted to
the construction of BPS-saturated supergravity solutions describing
various brane configurations including multiple branes, branes with
angular momentum, and other interesting features (see for example
\cite{Stelle:review, Youm:review} for reviews of some of the earlier
literature on this subject).

The most basic supersymmetric  black brane solutions of supergravity
are the bosonic R-R charged black $p$-brane solutions first described
in \cite{Horowitz-Strominger}.  These supergravity solutions correspond 
to the field
configurations around some number of coincident Dirichlet $p$-branes
in string theory.  A particularly simple example of these solutions is
the 10-dimensional supersymmetric extremal black hole in type IIA 
supergravity
which is charged under the R-R vector field.  This corresponds
to the supergravity field configuration around a D-particle (D0-brane)
of type IIA string theory.

Type IIA supergravity is related to 11-dimensional supergravity by
dimensional reduction on a single circle of radius $R =g l_s$, where
$g$ is the string coupling and $l_s$ is the string length.  Via this
reduction, the D0-brane of type IIA string theory is naturally related
to the KK particle associated with a supergraviton of momentum $1/R$ 
around the compact direction of the 11D theory.  The 11-dimensional 
supergraviton has 256 polarizations, and these lead to 256 polarization 
states of a D0-brane in the IIA theory. 

From the point of view of the D0-brane worldvolume theory,  the 
polarization states of the D0-brane arise from quantizing the fermions 
on the world-volume of the brane.  These worldvolume fermions  have 
couplings to the background fields of type IIA supergravity, and these 
couplings give rise to different physical properties for the different 
polarization states. In this paper we classify the various polarization 
states of the D0-brane and study their physical properties. In 
particular, we will see that different polarization states generate 
different long range fields, and determine the long range supergravity 
fields corresponding to an arbitrary D0-brane polarization state.

In a previous paper \cite{Mark-Wati-4} we found an explicit description 
of the
world-volume action of a system of multiple D0-branes in terms of the
component fields on the world-volume, up to terms describing the
linear coupling to a general supergravity background.  Restricting
this action to a single D0-brane located at the origin with zero
velocity, we find that the world-volume fermions couple to small
fluctuations in the background metric $h_{ \mu \nu}$, NS-NS 2-form
field $B_{\mu \nu}$ and R-R fields $C_{\mu}, C^{(3)}_{\mu \nu
\lambda}$ through the terms
\begin{equation}
S_{{\rm linear}} =  m_0 (h_{0i,j}(0) +C_{i,j}(0)) {i  \over 8} \theta 
\gamma^{ij} \theta + m_0 (B_{ij,k}(0) + C^{(3)}_{0ij,k}(0)) {i \over 16} 
\theta \gamma^{ijk} \theta +{\cal O} (\theta^4)
\label{eq:linear-action}
\end{equation}
where $m_0 = {1 \over g_s \sqrt{\alpha'}}$ is the D0-brane mass,  
$\theta$ is the 16-component spinor and $\gamma^i$ generate the
16-dimensional representation of the $SO(9)$ Clifford algebra.
These fermionic couplings for a single D0-brane were also previously
found by Morales, Scrucca and Serone \cite{mss}.
{}From (\ref{eq:linear-action}) it is clear that
certain choices of polarization for the D0-brane fermions will have
nonvanishing spin angular momentum (coupling to $h_{0i,j}$), magnetic 
D0-brane dipole moment (coupling to $C_{i,j}$),
D2-brane dipole moment (coupling to $C^{(3)}_{0ij,k}$) and magnetic 
H-dipole moment (coupling to $B_{ij,k}$).  Furthermore, it
is manifestly clear that the gyromagnetic ratio between the magnetic 
D0-brane dipole moment and spin angular momentum is always $g = 1$ and 
that similarly the D2-brane dipole moment and magnetic H-dipole moment 
have a ratio $g_{2H}= 1$.   It was shown in \cite{dlr} using 
supergravity 
methods that some D0-brane polarization states carry angular momentum 
and magnetic D0-dipole moment, and that the gyromagnetic ratio of 1 
implies that the spin-spin and dipole-dipole interactions precisely 
cancel so that the BPS zero force condition is satisfied.  As we show 
here, the agreement between the D2-dipole moment and magnetic H-dipole 
moment for any polarization state similarly guarantees that these two 
types of dipole-dipole interactions will also precisely cancel for pairs 
of D0-branes.

Since the different polarization states of a D0-brane couple differently 
to the supergravity background fields, they give rise to different 
supergravity solutions. These solutions are the ``superpartners'' 
\cite{Aichelburg-Embacher} of the 
bosonic D0-brane solution, obtained by acting 
on the bosonic solution 
with broken supersymmetry generators. Whereas the bosonic D0-brane 
solution has only dilaton, metric and RR one-form fields turned on, the 
solutions corresponding to arbitrary D0-brane polarizations generically 
have non-zero values for all the bosonic fields of type IIA supergravity 
including the RR three-form and NS-NS two form. The couplings we derive 
allow us to directly determine the long range supergravity fields 
corresponding to  an arbitrary polarization state of the D0-brane.

In this paper we will focus on the couplings between a single D0-brane
and the background supergravity fields.
In \cite{Mark-Wati-4}, we determined the analogous couplings for a
system of multiple D0-branes.  In the case of many branes, the
appearance of noncommuting matrices describing the space-time brane
configuration can lead to multipole moments of higher-dimensional
branes encoded in the bosonic matrix-valued brane coordinates.  
For example, a system of many D0-branes couples to the RR 3-form
through a term of the form
\[
C^{(3)}_{0ij, k}{\rm Tr}\; X^i[X^j, X^k]\;.
\]
It was pointed out by Myers \cite{Myers-dielectric} that as a
consequence of this coupling, the presence of a background 4-form
field strength $C^{(3)}_{0ij, k}$ produces a polarized D0-brane
configuration corresponding to a fuzzy D2-brane sphere.  In this paper
we show that such polarization can occur even for a single D0-brane,
since in the presence of a background 4-form, the couplings
(\ref{eq:couplings}) will break the degeneracy of the 256 states in
the multiplet associated with even a single D0-brane. We emphasize
that for a single D0-brane, the multipole moments we find are
fundamental moments, like the magnetic dipole moment of an electron,
whereas in Myers' dielectric effect, the configurations exhibiting
D2-brane dipole moments are spatially extended.

The paper is organized as follows. In section 2, we discuss the action 
for a single D0-brane in type IIA supergravity, and in particular, 
derive the linear couplings of the bulk fields to the worldvolume 
fermions of a static D0-brane. In addition to the quadratic terms 
found in previous work, we derive higher order terms with up to 8 
fermions which correspond to higher moments of various conserved 
quantities. We also provide an alternate derivation of the quadratic 
fermion terms from the $\kappa$-symmetric abelian D0-brane action, 
including an extension of the linearized result to the full non-linear 
D0-brane action to second order in the worldvolume fermion fields. In 
section 3 we discuss the supergravity solutions corresponding to the 
various states and in particular calculate the leading long range fields 
for an arbitrary polarization state. The expressions that we obtain 
depend on operators built out of the worldvolume fermions. In order to 
obtain the physical values of the fields for a given state, it is 
necessary to take the expectation values of these operator valued 
expressions in the state of interest.  In section 4, we describe an 
explicit representation of the D0-brane states and develop the machinery 
necessary to evaluate expectation values of any worldvolume fermion 
operators for arbitrary states. We also classify the states by providing 
an orthonormal basis of  eigenstates for a maximally commuting set of 
fermion operators.
Section 5 contains a discussion of the interaction between a pair of 
D0-branes, and in particular describes the complete cancellation of
dipole-dipole forces. We conclude in Section 6 with a brief discussion.

\section{D0-background couplings}

At low energies, a single D0-brane is described by simple 
non-relativistic quantum mechanics, with nine coordinates $X^i$ and dual 
momenta $P_i$ obeying the canonical commutation relations
\[
[X^i, P_j] = i \delta^i_j
\]
In addition, there are 16 fermionic operators $\theta_\alpha$ which obey 
\[
\{\theta_\alpha, \theta_\beta \} = \delta_{\alpha \beta}
\]
This is the 16-dimensional Clifford algebra, so we may represent the  
$\theta$'s by $2^{16/2} = 256$ dimensional gamma matrices, showing that 
the D0-brane has 256 independent states. 

The flat space Hamiltonian governing the low-energy D0-brane is  
the nonrelativistic action for a free particle
\[
H = {1 \over 2} P_i P_i
\]
which makes no reference to the fermions, so the 256 states are 
degenerate and apparently  indistinguishable.

However, as we will discuss in the next subsection, the fermionic 
operators $\theta$ do couple to background fields, even for a static 
D0-brane. 
Thus, turning on various background supergravity fields will break the 
degeneracy between the D0-brane states. Conversely, a D0-brane in flat 
space-time will generate a different set of supergravity fields 
depending on the state it is in. Thus, it is important to determine the 
couplings of worldvolume fermions to background fields in order to 
understand the different physical properties of the various polarization 
states.

\subsection{Couplings to background fields}

Ignoring the worldvolume fermions, the classical action for a 
single D0-brane in the presence of background type IIA supergravity 
fields is given in string frame by 
\be
\label{eq:D0ac}
S_{\rm bos} = - \mu_0 \int e^{-\phi} ds + \mu_0 \int C
\ee
where $\mu_0 = (\alpha')^{-{1\over 2}} $ is the D0-brane charge, $\phi$ 
is the dilaton field, 
$C$ is the RR one-form field, and $ds = \sqrt{ g_{\mu \nu} 
\dot{x^{\mu}} \dot{x^{\nu}}} d \tau$. Since the D0-brane mass is given 
by ${\mu_0 \over g_s} = \mu_0 e^{-\phi}$, this is essentially the action 
for a relativistic charged particle in a gravitational field. Together 
with the bulk type IIA supergravity action, this gives rise to the usual 
D0-brane supergravity solution, given in Einstein frame by 
\cite{Horowitz-Strominger}
\be
\label{eq:bosonic}
ds^2 = -H^{-{7 \over 8}}dt^2 + H^{1 \over 8}d\vec{x}^2, \qquad e^{\phi} 
= H^{3 \over 4}, \qquad C_0 = H^{-1} 
\ee
where $H$ is a harmonic function, 
\[
H = 1 + {60 \pi^3 g_s (\alpha')^{7 \over 2} \over r^7 } 
\]
In particular, we note that only the metric, dilaton, and RR one-form 
field are involved in the bosonic worldvolume action and supergravity 
solution. 

The story becomes more interesting with the inclusion of fermion fields. 
In a previous paper \cite{Mark-Wati-4}, we have derived leading terms in 
the 
D0-brane action coupling linearly to all type IIA supergravity 
background fields up to quadratic order in the fermion fields. 
At this order, the Lagrangian for a static D0-brane contains four terms 
involving fermions, namely (omitting an overall factor of $\mu_0$)
\bea
& &(\partial_i h_{0j} + \partial_i C_{j} ) {i \over 8} \theta 
\gamma^{ij} \theta 
+(\partial_i B_{jk} + \partial_i C^{(3)}_{0jk} ) {i \over 16} \theta 
\gamma^{ijk} \theta 
\label{eq:couplings}
\eea where $\gamma^{ij}$ and $\gamma^{ijk}$ are antisymmetrized
products of $16 \times 16$ symmetric Dirac matrices. 
These couplings agree with those found earlier in \cite{mss}.
The operator
$J^{ij} \equiv -{i \over 4} \theta \gamma^{ij} \theta$ represents the
fermionic part of the angular momentum (intrinsic spin) of the
D0-brane \cite{Kraus-spin} and has the expected coupling to the first
spatial derivative of the metric component $h_{0i}$. The coupling of
$J^{ij}$ to $\partial_i C_{j}$ indicates that D0-branes carry a
magnetic dipole moment (with respect to the RR one-form field)
proportional to the angular momentum, so a D0-brane has gyromagnetic
ratio 1. This fact, derived earlier as a property of D0-brane
supergravity solutions \cite{dlr}, is transparent from the form of
the worldvolume couplings.

   The second pair of couplings involves the operator $D^{ijk} \equiv {i
\over 4}\theta \gamma^{ijk} \theta$.  The coupling to $\partial_i
C^{(3)}_{0jk}$ indicates that certain polarization states of the
D0-brane carry dipole moments of D2-brane charge, while the coupling to
$\partial_i B_{jk}$ indicates that these states also have magnetic
H-dipole moments, where H is the field strength of the NS-NS two form
field.  Again, it is clear that the ratio of these dipole moments is 1 
in natural units.

  The couplings (\ref{eq:couplings}) were derived in
\cite{Mark-Wati-4} using the results of a Matrix theory calculation
and exploiting the relationship between Matrix theory and D0-branes in
type IIA string theory.  In \cite{mss}, these couplings were
determined by using the Green-Schwarz boundary state formalism to
compute the interaction between a pair of D0-branes.  Both of these
methods are somewhat indirect.  In principle, the extension of the
action (\ref{eq:D0ac}) to include fermions could also be determined
from the known $\kappa$-symmetric D0-brane action which is written
compactly using a $D=10$ superspace formalism. However, to derive the
couplings of worldvolume fermions to the background fields from this
action, it is necessary to expand the superfields in terms of the
component fields of type IIA supergravity and the worldvolume fields,
a non-trivial procedure which has not yet been carried out. In
subsection (\ref{sec:kappa}), we will perform this expansion
explicitly to quadratic order in the worldvolume fermions as a
check. These methods also yield the extension of (\ref{eq:couplings})
to the complete non-linear D0-brane action up to quadratic order in
$\theta$.

   Before discussing higher order terms, we note that there is a very
natural understanding of the ratios $g=1$ and $g_{2H}=1$ based on the
11-dimensional origin of the D0-brane action. In the Matrix theory
action describing a single graviton with zero transverse momentum in
DLCQ supergravity \cite{Mark-Wati-3}, there are two couplings between
background $D=11$ supergravity fields and bilinears of the Matrix
theory fermions (which describe the supergraviton polarizations),
proportional to $\partial_j h_{+i} \theta \gamma^{ij} \theta$ and
$\partial_k A_{+ij} \theta \gamma^{ijk} \theta$. Recalling the
relations (to linear order)
\[
C^{IIA}_i = h^{11}_{10 \; i}, \qquad h^{IIA}_{0i} = h^{11}_{0i}, \qquad 
B^{IIA}_{ij} = A^{11}_{10\;ij}, \qquad C^{IIA}_{0ij} = A^{11}_{0ij}
\]
we see that the first two couplings in (\ref{eq:couplings}) arise from 
the single term $\partial_j h_{+i} \theta \gamma^{ij} \theta$ term in 
eleven dimensions, leading to $g=1$, while the second pair of couplings 
in (\ref{eq:couplings}) arises from the single term $\partial_k A_{+ij} 
\theta \gamma^{ijk} \theta$ term in eleven dimensions, leading to 
$g_{2H} = 1$. For a more precise discussion of the relationship between 
the Matrix theory action and D0-brane action, see \cite{Mark-Wati-4}.

\subsection{Couplings with more than two fermions}

Additional couplings between fermions and background fields exist with 
four, six, eight and possibly up to sixteen fermions. One way to derive 
these would be to extend 
the Matrix theory calculation of \cite{Mark-Wati-3} to higher orders in 
$1/r$ and to 
higher order in the fermion fields. The required calculation would be 
rather tedious, but we will be able to deduce some of these couplings 
here based on other 
considerations.

In general, we may write the linear couplings of background fields to 
the worldvolume fields in the D0-brane action as
\bea
&&\int dt  \sum_{n=0}^{\infty} {1 
\over n!} \left[ 
\frac{1}{2}
(\partial_{k_1}\cdots 
\partial_{k_n} h_{\mu \nu}) \; I_h^{\mu \nu (k_1 \cdots k_n)}
+ (\partial_{k_1}\cdots \partial_{k_n} \phi) 
\; I_{\phi}^{(k_1 \cdots k_n)}\right.\label{eq:IIA-general}\\ 
& & \hspace{1in}+ (\partial_{k_1}\cdots 
\partial_{k_n} C_{\mu }) \; I_0^{\mu (k_1 \cdots k_n)}
+
 (\partial_{k_1}\cdots 
\partial_{k_n} \tilde{C}_{\mu \nu \lambda \rho \sigma \tau \zeta }) 
\; I_6^{\mu \nu \lambda
\rho \sigma \tau \zeta (k_1 \cdots k_n)}
\nonumber\\
& & \hspace{1in}+
 (\partial_{k_1}\cdots \partial_{k_n} 
B_{\mu \nu}) \; I_s^{\mu \nu (k_1 \cdots k_n)} 
+ 
(\partial_{k_1}\cdots \partial_{k_n} 
\tilde{B}_{\mu \nu \lambda \rho \sigma \tau}) \; I_5^{\mu \nu \lambda
\rho \sigma \tau
 (k_1 \cdots k_n)} \nonumber\\
& &\hspace{1in} \left.+ 
(\partial_{k_1}\cdots 
\partial_{k_n} C^{(3)}_{\mu \nu \lambda }) \; I_2^{\mu \nu \lambda (k_1 
\cdots 
k_n)}
+
 (\partial_{k_1}\cdots 
\partial_{k_n} \tilde{C}^{(3)}_{\mu \nu \lambda \rho \sigma }) 
\; I_4^{\mu \nu \lambda \rho \sigma (k_1 \cdots 
k_n)}\right] \nonumber
\eea
Here, $I_h^{\mu \nu (k_1 \cdots k_n)}$, which couples to $n$ derivatives 
of the metric, represents the $n$-th spatial moment of the stress energy 
tensor for a D0-brane, while the remaining operators $I$ represent 
moments of various other currents. 

We would like to derive expressions for the moments $I$ for a static 
D0-brane at the 
origin. The couplings for a moving D0-brane (velocity dependent terms) 
could then be deduced by performing a boost on this action. In the 
static 
case, the expressions $I$ are built only out of the fermion 
fields $\theta$ as well as $\gamma$-matrices, since we have 
$X^i=\dot{X}^i=0$. There are various constraints that allow us to 
determine leading terms in the various couplings:
\\ \\
{\bf Symmetries}\\
 Of course, the D0-brane action should have $SO(9)$ rotational 
invariance. This requires that all spatial indices in the action be 
contracted with other spatial indices.
\\ \\
{\bf Consistency with T-duality} \\
The D0-brane action is related by T-duality to all of the
higher-dimensional Dp-brane actions. For a given $p$, the Dp-brane
action must  
have $p+1$ dimensional Lorentz invariance as well as $SO(9-p)$ 
rotational invariance. The requirement of these symmetries in the dual 
actions places further constraints on the form of the D0-brane action. 
In particular, it requires that all operators $I$ may be written as the 
dimensional reduction of $D=10$ Lorentz invariant objects, just as the 
low-energy flat-space part of the action is the dimensional reduction of 
$D=10$ SYM theory. For example, under a T-duality of all nine spatial 
directions, a weak background metric transforms as $h^{IIA}_{ij} \to 
-h^{IIB}_{ij}$. Thus, the operator coupling to $h_{ij}$ in the D0-brane 
action is the T-dual version of the operator coupling to $-h^{IIB}_{ij}$ 
in the D9-brane action. In this case, T-duality acts on the worldvolume 
operator simply by dimensional reduction, thus, we conclude that the 
operator coupling to $h_{ij}$ in the D0-brane action must be the 
dimensional reduction of a $D=10$ Lorentz covariant object.

The requirement of $D=10$ Lorentz covariance suggests that the D0-brane 
action should be written most compactly using $D=10$ notation for the 
fermions, i.e. 32-component Majorana-Weyl spinors and $32 \times 32$ 
Dirac matrices. Taking into account the Weyl property of the spinors as 
well as their anticommutation relation, it is evident that all 
non-vanishing expressions may be built from fermion bilinears with 
three-index antisymmetric products of $\Gamma$-matrices
\be
\label{eq:bil}
\bar{\Theta} \Gamma^{abc} \Theta \;
\ee
where $a, b, c \in\{0, \ldots, 9\}$.

To summarize, the purely fermionic parts of the operators $I$
should be constructed out of $D=10$ fermion bilinears (\ref{eq:bil}). 
The operators should display $D=10$ Lorentz covariance, thus all indices 
apart from those contracted with background fields or derivatives must 
be contracted with each other and summed from 0 to 9.\footnote{Note 
that the RR couplings in Dp-brane actions include a $p+1$ index 
antisymmetric epsilon tensor when expressed in terms of components. 
Thus, 
for each coupling $C I_C$ in the D0-brane action, either the RR field 
$C$ or the tensor $I$ must contain a single free 0 index which 
``contracts'' with the trivial $\epsilon$ tensor $\epsilon^0$.} In 
constructing 
operators with more than two fermions, it is necessary to take into 
account various Fierz identities listed in the appendix.

In many cases, these requirements completely determine the operators $I$ 
up to a coefficient. 
\\ \\
{\bf Relation to $N=4$ $D=4$ chiral operators}
\\ \\
In some cases, the fermionic terms in the moments may be determined from 
the known expressions for the purely bosonic terms using supersymmetry 
and the AdS/CFT correspondence. For a given field $\phi$, the lowest 
dimension operator coupling to $\phi$ in the D3-brane action is 
essentially the chiral operator of ${\cal N} = 4$ SYM theory related to 
the particle $\phi$ via the AdS/CFT correspondence.\footnote{This is 
most clear for minimally coupled scalars which are described by the same 
fields everywhere. For a discussion of the more general case, see 
\cite{Mark-Wati-5}} 
These chiral operators may be obtained by acting with supersymmetry 
generators on chiral primary operators of the form 
\[
\str(X^{i_1} \cdots X^{i_n}) -  \{ {\rm traces} \}
\]
By determining the correct combination of supersymmetry generators that 
reproduce the known bosonic terms in an operator, the purely fermionic 
terms may be easily deduced. This method was used in \cite{ktv} to 
determine the four 
fermion terms in the operator coupling to the Einstein frame dilaton in 
the D3-brane action. Such purely fermionic terms in the D3-brane action 
may be T-dualized to give the desired fermionic couplings in the 
D0-brane action including coefficients.

\subsection{Results for the D0-brane action to linear order in 
background fields}

Assuming the validity of the considerations above, we find the following 
terms in the action for a static D0-brane in the presence of arbitrary 
type IIA supergravity 
fields. We omit overall factors of D0-brane mass/charge.
\\ \\
{\bf zero fermion terms: charges}
\\ 
\[
S_0 = \int d\sigma ({3 \over 4}\phi + {1 \over 2} h_{00} + C_0) 
\]
These terms, written in Einstein frame, indicate that the D0-brane is 
massive and charged under the RR one-form field. The coupling to the 
dilaton field arises because the D0-brane mass is inversely proportional 
to the string coupling.
\\ \\
\noindent
{\bf two fermion terms: dipole moments}
\\ 
\bea
S_2 &=& -{i \over 8} \partial_j h_{0i} \bar{\Theta} \Gamma^{0ij} \Theta 
- {i \over 8} \partial_j C_i \bar{\Theta} \Gamma^{0ij} \Theta 
\nonumber\\
& &+ {i \over 16} \partial_k B_{ij} \bar{\Theta} \Gamma^{ijk} \Theta + 
{i \over 16} \partial_k C^{(3)}_{0ij}  \bar{\Theta} \Gamma^{ijk} \Theta
\label{eq:couplings2}
\eea
These are exactly the terms (\ref{eq:couplings}) discussed above, 
written now in $D=10$ 
notation. They indicate that certain polarization states may carry 
angular momentum, RR one-form magnetic dipole moment, D2-brane dipole 
moment and magnetic H-dipole moment.
\\ \\
{\bf four fermion terms: quadrupole moments}\footnote{In expressions 
containing terms with four or more fermions, there is an ordering 
ambiguity that arises when promoting a classical action to an operator 
expression, since the $\Theta$ operators obey a non-trivial 
anticommutation relation. The fermion operators described in this 
section are related to chiral operators of $N=4$ SYM theory which are 
obtained by acting with supersymmetry transformations on symmetrized 
traces of bosonic fields. This suggests that the correct resolution of 
the ordering ambiguity is an antisymmetrization of the $\Theta$s in each 
of the expressions here.}
\\ 
\beas
S_4 &=& {1 \over 384} \partial_k \partial_l h_{00} \bar{\Theta} 
\Gamma^{0ak} \Theta \; \bar{\Theta} \Gamma^{0al} \Theta - {1 \over 384}  
\partial_k \partial_l  h_{ij} \bar{\Theta} \Gamma^{aik} \Theta \; 
\bar{\Theta} \Gamma^{ajl} \Theta \\
& & -{1 \over 192} \partial_k \partial_l B_{0i}\bar{\Theta} \Gamma^{a0k} 
\Theta \; \bar{\Theta} \Gamma^{ail} \Theta + {1 \over 32} \partial_k 
\partial_l C^{(3)}_{ijk}\bar{\Theta} \Gamma^{0il} \Theta \; \bar{\Theta} 
\Gamma^{jkm} \Theta \\
& & + {1 \over 128} \partial_m \partial_n C^{(5)}_{0ijkl}\bar{\Theta} 
\Gamma^{mij} \Theta \; \bar{\Theta} \Gamma^{nkl} \Theta
\eeas
The first two terms here indicate quadrupole moments for the $00$ and 
$ij$ components of the stress-energy tensor. The third term indicates a 
quadrupole moment of string electric charge ($B_{0i}$). The RR couplings 
both correspond to a magnetic quadrupole moment of D2-brane charge or 
equivalently 
an electric quadrupole moment of D4-brane charge. 
These quadrupole moments were previously found in \cite{mss}.
\\ \\
{\bf six fermion terms: octupole moments}
\\ 
\beas   
S_6 &=& {i \over 5760} \partial_j \partial_k \partial_l h_{0i}  
\bar{\Theta} \Gamma^{a0k} \Theta \; \bar{\Theta} \Gamma^{bil} \Theta \;  
\bar{\Theta} \Gamma^{abj} \Theta  \\
& & + { i \over 11520} \partial_k \partial_l \partial_m B_{ij}  
\bar{\Theta} \Gamma^{aik} \Theta \; \bar{\Theta} \Gamma^{bjl} \Theta \;  
\bar{\Theta} \Gamma^{abm} \Theta \\
& & +  c_1 \;  \partial_p \partial_q \partial_r C^{(5)}_{jklmn}  
\bar{\Theta} \Gamma^{0jp} \Theta \; \bar{\Theta} \Gamma^{klq} \Theta \;  
\bar{\Theta} \Gamma^{mnr} \Theta \\ 
& & + { c_1 \over 6} \partial_p \partial_q \partial_r C^{(7)}_{0ijklmn}  
\bar{\Theta} \Gamma^{ijp} \Theta \; \bar{\Theta} \Gamma^{klq} \Theta \;  
\bar{\Theta} \Gamma^{mnr} \Theta 
\eeas
Here, $c_1$ is a numerical constant that we have not determined. These 
terms indicate an octupole moment of the $0i$ component of the stress 
energy tensor, a magnetic H-octupole moment, an octupole moment of 
D6-brane charge and an octupole moment of D2-brane charge (or magnetic 
D4-brane charge).
\\ \\ 
{\bf eight fermion terms: 16-pole moments}
\\ \\
At this order, we only mention one additional term, 
\[
S_8 = c_2 \;  \partial_p \partial_q \partial_r \partial_s 
C^{(7)}_{ijklmnp}  \bar{\Theta} \Gamma^{0iq} \Theta \; \bar{\Theta} 
\Gamma^{jkr} \Theta \;  \bar{\Theta} \Gamma^{lms} \Theta \bar{\Theta} 
\Gamma^{npt} \Theta
\]
which represents a 16-pole moment of magnetic D6-brane charge (or 
electric D0-brane charge).

The couplings we have listed are those related to nonvanishing terms in 
the action for Matrix theory in a $D=11$ supergravity background. In 
particular, we have derived the leading terms coupling linearly to all 
the background supergravity fields. We expect additional terms that 
vanish in the Seiberg-Sen limit, possibly up to sixteen fermion terms. 
The anticommutation relations for the fermions ensure that no terms with 
more than sixteen fermions exist, so there are certainly only a finite 
number of moments carried by a single D0-brane, as we would expect for a 
point particle. 

\subsection{Fermion couplings from the superspace action.}
\label{sec:kappa}

We have derived the fermionic couplings in the D0-brane action by 
somewhat indirect methods, exploiting various symmetries and dualities 
as well as connections with Matrix theory and the AdS/CFT 
correspondence. In principle, we could also have derived all of these 
terms directly using the known $\kappa$-symmetric superspace action for 
a D0-brane. As an demonstration of this approach and a check of the 
results in the previous section, we now calculate the two-fermion terms 
directly from the superspace action. 

The $\kappa$-symmetric action for a D0-brane is given by 
\cite{aps1,aps2,cgnsw,Bergshoeff-Townsend} 
\[
S = -\mu_0 \int dt\left( e^{-{3 \over 4} \Phi }\sqrt{-\Pi _{0}^{r}\Pi 
_{0}^{s}\eta_{rs}}-\partial _{0}Z^{M}B_{M}\right) \; ,
\]
 where \( \Pi _{0}^{r}=\partial _{0}Z^{M}E_{M}^{r} \) 
is the pullback of the supervielbein to the world-line of the 
D0-brane\footnote{Type IIA superspace is defined by coordinates $Z^M$, 
$M = 
\{ 
\mu, a \}$ , where $Z^\mu \equiv X^\mu$ are the 10 bosonic coordinates 
and  $Z^a \equiv \Theta^a$ is a pair of Majorana-Weyl spinors of 
opposite chirality combined into a single Majorana spinor. In this 
section, we use indices $M, \mu, \alpha$ (and similar) to represent 
spacetime 
superfield, bosonic, and fermionic indices respectively and $A, r, 
a$ (and similar) to represent tangent space superfield, bosonic and 
fermionic indices respectively.} In order to find the coupling of the 
D0-brane to the component background fields, we need to expand the 
superfields out in terms of the component fields of type IIA 
supergravity. To do this, we perform an order by order expansion in the 
fermionic superspace coordinates \( \Theta  \), in a method known as 
gauge completion. A similar expansion was carried out to 
second order for eleven-dimensional supergravity in \cite{dpp}, and our 
analysis will mirror that in their paper closely. The expansion of type 
IIA superspace fields in terms of component fields has also been 
discussed in \cite{clps}.
An alternative approach to expanding the vielbein $E^r_M$ in terms of
fermionic component fields was taken in \cite{gks}, where the authors
performed a superspace analog of the Riemann normal coordinate 
expansion.

The method of gauge completion involves reconciling the component  
forms of the supersymmetry transformations of the fields with superspace 
diffeomorphisms at each order in \( \Theta  \). In this way, starting 
from order \( \Theta ^{0} \) we can build up the superfields in terms of 
their components in an expansion in \( \Theta  \). 

The supersymmetry transformations of the relevant fields in type IIA 
supergravity are listed in appendix B. A given component field 
supersymmetry transformation corresponds to some combination of a 
superspace diffeomorphism, a Lorentz transformation and a gauge 
transformation, whose action on the fields appearing in the D0-brane 
action is
\begin{eqnarray}
\delta \Phi & = & \Xi^M \partial_M \Phi \nonumber \\
\delta E_{M}^{A} & = & \partial _{M}\Xi ^{N}E_{N}^{A}+\Xi ^{N} 
\partial_{N}E_{M}^{A} + \Lambda^A {}_B E_M^B \label{eq:supertrans}\\
\delta B_{M} & = & \partial _{M}\Xi ^{N}B_{N}+\Xi ^{N}\partial _{N}B_{M} 
+ \partial_M \Omega \; .\nonumber
\end{eqnarray}
Here, \( \Xi^M  \) is the superdiffeomorphism parameter, $\Omega$ is the 
gauge transformation parameter for $B$, and $\Lambda^A {}_B$ is the 
tangent space Lorentz transformation (with nonvanishing components 
$\Lambda^r {}_s$ and $\Lambda^a {}_b = {1 \over 4} \Lambda^{rs} 
(\Gamma_{rs})^a {}_b$). Each of these superspace parameters is 
some function of the component fields, the component field 
transformation parameters and the superspace coordinate $\Theta$. 

At zeroth order in \( \Theta  \), we identify the components of the 
superfields and the transformation parameters as 
\[
\ba{ll}
E_{\mu }^{r}  =  e_{\mu }^{r} \qquad \qquad  & B_{\mu }  =   C_{\mu }\\
E_{\mu }^{a }  =  \psi _{\mu }^{a } \qquad \qquad& B_{\alpha}  =  
0\\
E_{\alpha}^{r}  =  0 \qquad \qquad& \Xi ^{\mu }  =  \xi^\mu\\
E_{a}^{\alpha }  =  \delta _{a}^{\alpha } \qquad \qquad& \Xi ^{\alpha }  
= \epsilon ^{\alpha }\\ 
\Phi = \phi \qquad \qquad  & \Omega = \omega \\
\Lambda^{rs} = \lambda^{rs} & 
\ea
\]
where $\epsilon_\alpha$, $\xi^\mu$, $\lambda^{rs}$, and $\omega$ 
parameterize the usual component field supersymmetry transformations, 
diffeomorphisms, local Lorentz transformations, and gauge 
transformations on the RR one-form respectively.

We begin by finding the transformation parameters \( \Xi, \Lambda, 
\Omega  \) to first order in \( \Theta  \) by 
demanding that the $\Theta^0$ terms in the commutator of superspace 
transformations applied to a superfield match with the commutator of the 
component field transformations on the appropriate component field. 
Using this, we find to order $\Theta$
\begin{eqnarray*}
\Xi^{\mu } & = & \xi^\mu  - {i \over 2} \bar{\Theta }\gamma ^{\mu 
}\epsilon \\
\Xi^{\alpha } & = & \epsilon ^{\alpha }\\
\Omega & = & \omega + {i \over 2} \bar{\Theta} \gamma^\rho \epsilon 
C_\rho + {i \over 2} e^{-{3 \over 4} \phi} \bar{\Theta} \Gamma^{11} 
\epsilon\\
\Lambda^{pq} &=& \lambda^{pq} - {i \over 2} \bar{\Theta} \gamma^\mu 
\epsilon \; 
\omega_\mu 
{}^{pq} -  {i \over 64} e^{{3 \over 4} \phi} \bar{\Theta} \gamma^{pq 
\nu \lambda} \Gamma^{11} \epsilon F_{\nu \lambda} - {7i \over 32} e^{{3 
\over 4} \phi} \bar{\Theta} \Gamma^{11} \epsilon F^{pq}\\
& &+{i \over 96} e^{-{1 \over 2} \phi} \bar{\Theta} \gamma^{pq \nu 
\lambda 
\sigma} \Gamma^{11} \epsilon H_{\nu \lambda \sigma} + {3i \over 16} 
e^{-{1 \over 2} \phi} \bar{\Theta} \gamma_{\lambda} \Gamma^{11} \epsilon 
H^{pq \lambda} \\
& &+{i \over 256} e^{{1 \over 4} \phi} \bar{\Theta} \gamma^{pq \nu 
\lambda 
\sigma \tau} \epsilon F'_{\nu \lambda \sigma \tau} +{5i \over 64} e^{{1 
\over 4} \phi} \bar{\Theta} \gamma_{\sigma \tau} \epsilon F'^{ab  \sigma 
\tau} 
\end{eqnarray*}
Here, and in the following, we set to zero all background fermion 
fields, except where they are important in determining the desired terms 
in the final action.

To determine the superspace fields at first order in \( \Theta  \), we 
write down the order $\Theta^0$ terms in the superspace transformations 
(\ref{eq:supertrans}) and demand that the resulting equations are 
consistent with the component field supersymmetry transformations. We 
find, up to ambiguities that can be removed by gauge transformations, 
\begin{eqnarray*}
\Phi &=& \phi + i \sqrt{2} \bar{\Theta} \Gamma^{11} \lambda\\
E_{\mu }^{r} & = & e_{\mu }^{r} + i \bar{\Theta }\Gamma^{r}\psi _{\mu 
}\\
E^{r}_{\alpha } & = & {i \over 2} (\bar{\Theta }\Gamma ^{r})_{\alpha 
}\\
B_{\mu } & = & C_{\mu }+ {3i\sqrt{2} \over 4} e^{- {3 \over 4} \phi} 
\bar{\Theta }\gamma _{\mu }\lambda 
 + i e^{- {3 \over 4} \phi} \bar{\Theta}\Gamma^{11}\psi_{\mu }\\
B_{\alpha } & = & -{i \over 2} e^{- {3 \over 4} \phi}(\bar{\Theta 
}\Gamma 
^{11})_{\alpha }
\end{eqnarray*}
Note that the order $\Theta$ terms in the bosonic components of the 
superspace fields are precisely the supersymmetry 
variations of the order $\Theta^0$ terms in these fields with 
supersymmetry variation parameter $\Theta$, as we would expect. 

The calculation of order $\Theta^2$ terms in the superspace fields and 
transformation parameters proceeds in the same way as for order 
$\Theta$.
In order to evaluate the order $\Theta^2$ terms in the D0-brane action, 
it is only necessary to determine \( E_{\mu }^{r} \), \( B_{\mu } \), 
and $\Phi$ to order \( \Theta ^{2} \), and these may be deduced 
immediately 
by demanding that the order $\Theta$ terms in (\ref{eq:supertrans}) 
agree with the component field supersymmetry transformations. The 
resulting expressions are somewhat complicated, but may be simplified 
considerably by fixing $\kappa$-symmetry. We set half of the 32 
components of $\Theta$ to zero making the gauge choice ${1 \over 2}(1 - 
\Gamma^{11}) \Theta = \Theta$. We have made this choice so that the 
remaining spinor $\Theta$ may be identified with the worldvolume 
$\Theta$ appearing in earlier expressions for the D0-brane action. The 
resulting $\Theta^2$ components are then given by\footnote{It is 
interesting to note that the $\Theta^2$ terms in these bosonic 
components may be obtained more simply by the observation that they are 
exactly the anticommutator of two supersymmetry variations on the order 
$\Theta^0$ terms, that is
\[
A|_{\Theta^2} = {1 \over 2} \{ \delta_{\epsilon_1}, \delta_{\epsilon_2} 
\} a |_{\epsilon_1 = \epsilon_2 = \theta} 
\] where $a$ is the component field $a = A|_{\Theta^0}$. Thus, we could 
have obtained the desired expressions in one step directly from the 
supersymmetry transformations.}
\beas
\Phi|_{\Theta^{2}} &=& {i \over 48} e^{-{1 \over 2}\phi} \bar{\Theta} 
\gamma^{\mu \nu \lambda} \Theta H_{\mu \nu \lambda}\\
B_{\mu }|_{\Theta ^{2}} &=& -{i \over 16}\bar{\Theta }\gamma_\mu {}^{\nu 
\lambda } \Theta F_{\nu \lambda } - {i \over 48} e^{-{1 \over 2} \phi} 
\bar{\Theta }\gamma^{ \nu \lambda \rho} \Theta F'_{\mu \nu \lambda 
\rho}\\
E_{\mu }^{r}|_{\Theta ^{2}} &=& {i \over 8} \bar{\Theta} 
\gamma^{rpq}\Theta \omega_{\mu pq} +{i \over 64}e^{-{1 \over 2} 
\phi}\bar{\Theta }\gamma_\mu {}^{\nu \lambda } \Theta H^r{}_{\nu \lambda 
} \\
& & +{3i \over 64} e^{-{1 \over 2} \phi} \bar{\Theta }\gamma^{r \nu 
\lambda} \Theta H_{\mu \nu \lambda} - {i \over 192} e_\mu^r  \bar{\Theta 
}\gamma^{ \nu \lambda \rho} \Theta H_{ \nu \lambda \rho}
\eeas
In terms of these expressions, the complete non-linear action for a 
D0-brane to order $\Theta^2$ is then given by\footnote{Note that we may 
set to zero any terms involving derivatives on $\Theta$ since the 
worldvolume fermions are non-dynamical.}
\beas
S &=& -\mu_0 \int d \tau \; e^{- {3 \over 4} \phi} (1 - {3 \over 4} \Phi 
|_{\Theta^2} + \dots)\sqrt{-(g_{\mu \nu} + 2 e_{\mu 
r}E_{\nu}^r|_{\Theta^2} + \dots )\dot{x}^\mu \dot{x}^\nu}\\
& & +\mu_0 \int d \tau \; (C_\mu + B_\mu |_{\Theta^2} + 
\dots)\dot{x}^\mu
\eeas
where dots indicate terms at fourth or higher order in $\Theta$. 
Choosing the static gauge $X^0 = \tau$ and taking the weak field 
approximation (keeping only terms linear in the background fields), the 
velocity independent terms reduce to
\[
S_{\rm weak} = -{i \over 8} (\partial_j h_{0i} + \partial_j 
C_i)\bar{\Theta} \Gamma^{0ij} \Theta  + {i \over 16} (\partial_k C_{0ij} 
+ \partial_kB_{ij}) \bar{\Theta} \Gamma^{ijk} \Theta
\]
These are in agreement with the results from the previous section, 
providing an alternate derivation and a check of the results
previously discussed, as well as 
an extension to the complete non-linear action to order $\Theta^2$.  In 
principle, the expansion of the superspace fields in terms of components 
could be extended to higher orders in $\Theta$ to reproduce and check 
the higher order terms that we have derived, but we will not attempt 
this here.  For an expansion to high order in $\Theta$, it might be
that the superspace normal coordinate approach of \cite{gks} would
lead more efficiently to results which could be compared with those of
the previous subsection.

\section{D0-brane supergravity solutions}

An important physical effect of the fermion couplings derived in the 
previous section is that D0-branes in different polarization states will 
produce different long range supergravity fields. This phenomenon has 
been discussed previously in \cite{dlr,mss} (and also \cite{bktw} in the 
context of the 
M2-brane). In those papers, it was explained that the supergravity 
solution corresponding to an arbitrary polarization state could be 
obtained by acting iteratively with broken supersymmetry generators on 
the fields of the usual bosonic supergravity solution.
The result is a supergravity solution with fields depending on the 
fermionic supersymmetry transformation parameter. In \cite{bktw}, it was 
pointed 
out that this fermionic parameter should be identified with quantized 
fermion zero-mode operators, which from the worldvolume point of view 
are the 16 
non-dynamical  worldvolume spinors. The classical supergravity solution 
corresponding to a given state can then be evaluated by taking the 
expectation value of these operator valued supergravity fields in the 
state of interest. 

The fermionic couplings derived in the previous section give a direct 
understanding of how different polarization states lead to different 
supergravity solutions. With our results, it is possible to directly 
read off the long range supergravity fields corresponding to a given 
state. In the next subsection, we give the leading long range behavior 
for each field in type IIA supergravity as a function of the fermionic 
operators $\theta$. 

One important result, already clear from the results of the previous
section is that the RR three-form field $C^{(3)}_{\mu \nu \lambda}$
(as well as its dual $C^{(5)}$) and the NS-NS two form field $B_{\mu
\nu}$ are generically non-zero in D0-brane supergravity solutions. In
\cite{dlr}, it was assumed that these fields were not relevant to the
D0-brane solutions, so the D2-brane dipole moments and magnetic
H-dipole moments of D0-brane polarization states were not found. In
subsection \ref{sec:superpartner}, we rederive these moments using the
methods of \cite{dlr} by dropping the assumption that $C^{(3)}$ and
$B$ vanish.

\subsection{Long range fields}

In this section we write down the leading long range supergravity fields 
corresponding to  an arbitrary D0-brane polarization state. The long 
range fields are determined by the couplings derived in section 2 as 
well as type IIA bulk supergravity action, given to quadratic order in 
the Einstein frame by
\be
\label{eq:lin}
S_{IIA} = {1 \over 2 \kappa^2} \int d^{10} x \left\{ R - {1 \over 2} 
\partial_\mu \phi \partial^\mu \phi - {1 \over 12} |dB|^2 - {1 \over 4} 
|dC^{(1)}|^2 - {1 \over 48} |dC^{(3)}|^2 \right\}
\ee
Choosing the standard gauges $\partial^\mu (h_{\mu \nu} - (1/2) 
\eta_{\mu \nu} h^\lambda{}_\lambda) = 0$, $\partial^\mu C^{(0)}_\mu = 
0$, $\partial^\mu  B_{\mu \nu} = 0$ and $\partial^\mu C^{(3)}_{\mu \nu 
\lambda} = 0$, we may use the equations of motion derived from 
(\ref{eq:lin}) and (\ref{eq:IIA-general}) to determine the following 
long range fields, expressed in terms of the multipole moment operators 
$I$ appearing in (\ref{eq:IIA-general}).
\beas
h^{\alpha \beta} &=& \sum_n (-1)^n{15 \kappa^2 \over 16 \pi^4 n!} \left( 
I_h^{\alpha \beta(i_1 \cdots i_n)} - {1 \over 8} \eta^{\alpha \beta} 
(I_h)_\mu {}^{\mu (i_1 \cdots i_n)} \right) \partial_{i_1} \cdots 
\partial_{i_n} \left\{ {1 \over r^7} \right\} \\
\phi &=& \sum_n(-1)^n {15 \kappa^2 \over 16 \pi^4 n!} I_\phi^{(i_1 
\cdots 
i_n)} \partial_{i_1} \cdots \partial_{i_n} \left\{ {1 \over r^7} 
\right\}\\
D^{\mu_1 \cdots \mu_k} &=& \sum_n (-1)^n{15 \kappa^2 k! \over 16  \pi^4 
n!} I_C^{\mu_1 \cdots \mu_k (i_1 \cdots i_n)} \partial_{i_1} \cdots 
\partial_{i_n} \left\{ {1 \over r^7} \right\}
\eeas
Here, $D$ stands for any of the form fields $B$, $C^{(1)}$,  $C^{(3)}$, 
$C^{(5)}$ or $C^{(7)}$. In terms of the string theory parameters, the 
gravitational coupling is given by $\kappa^2 = 2^6 \pi^7 g_s^2 
(\alpha')^4$. 

The expressions for the $I$s may be determined by comparing the 
couplings derived in section 2 with the general expression 
(\ref{eq:IIA-general}). Since 
we are dealing only with the linearized theory, the long range fields 
generated by each of the couplings in section 2 may be calculated 
separately and combined where necessary by superposition. As an example, 
we find all long range fields at order ${1 \over r^8}$. 
These fields are associated with dipole moments and arise from the 
couplings (\ref{eq:couplings2}) quadratic in 
fermions. They are given by
\bea
h_{0i} &=& -{15 \over 2} \pi^3 i (\alpha')^{7 \over 2} \bar{\Theta} 
\Gamma^{0ij} \Theta 
\partial_j \left\{ {1 \over r^7} \right\} \nonumber\\
C_i &=& {15 \over 2} \pi^3 i (\alpha')^{7 \over 2} \bar{\Theta} 
\Gamma^{0ij} \Theta \partial_j \left\{ {1 \over r^7} \right\} 
\nonumber\\
C_{0ij} &=&  {15 \over 2} \pi^3 i (\alpha')^{7 \over 2} \bar{\Theta} 
\Gamma^{ijk} \Theta \partial_k \left\{ {1 \over r^7} \right\} 
\label{eq:suglin}\\
B_{ij} &=& - {15 \over 2} \pi^3 i (\alpha')^{7 \over 2} \bar{\Theta} 
\Gamma^{ijk} \Theta \partial_k \left\{ {1 \over r^7} \right\} \nonumber
\eea
These order ${1\over r^8}$ terms were also found in \cite{mss}.  The
corrections at orders ${1\over r^9}$ and higher may be similarly 
determined from the couplings of section 2 involving four or more 
fermions. Note that these fields depend on the worldvolume fermion 
operators $\Theta$. To determine the numerical values of a field for a 
given polarization state, we simply replace these expressions by their 
expectation value in the particular state. In section 4, we will give an 
explicit representation of the polarization states and develop the 
methods to compute these expectation values.

\subsection{Fermionic couplings from superpartner solutions}
\label{sec:superpartner}

In this section, we use the methods of \cite{dlr} to provide an 
alternate derivation that 
the supergravity solution corresponding to an arbitrary D0-brane 
polarization state has long range fields consistent with the new 
couplings that we have derived. In particular, by applying broken 
supersymmetry generators to the bosonic D0-brane solution, we will 
generate the ``superpartner'' solutions whose fields depend on the 
supersymmetry variation parameter. Relating this spinor parameter 
with the non-dynamical worldvolume fermions, we will show that the new 
subleading (order $1/r^8$) long range fields in the superpartner 
solutions match with the results of the previous subsection and 
therefore correspond precisely to the quadratic fermion couplings we 
have derived. In particular, we will see the presence of a 
non-zero RR three-form field corresponding to a D2-brane dipole directly 
from the supergravity solutions. This will provide an independent check 
of these results.   

We begin with the bosonic D0-brane solution (\ref{eq:bosonic}) which has 
a non-vanishing metric, dilaton, and RR one-form field.

If we let $\Phi$ denote the fields of the bosonic D0-brane solution, the 
superpartner solution is given by $\tilde{\Phi} = e^{\epsilon_\alpha 
Q_\alpha} \Phi$, where $Q_\alpha$ are the 16 broken supersymmetry 
generators. This is analogous to obtaining a spatially translated 
solution by exponentiating the broken translation generator, 
$\Phi(\vec{x} + \delta \vec{x}) = e^{\delta x^i P_i} \Phi$, however in 
our case, the exponential series must truncate at order 16 since 
$\epsilon$ is a 
Grassman quantity with 16 independent components. Each application of a 
pair of $Q$'s results in an additional derivative on the original 
fields, so the leading effects of the $\epsilon^{2n}$ terms in the long 
range fields will be at order ${1\over r^{7+n}}$. Thus, in order to 
compare the dipole fields of the supergravity solution with those 
predicted by the quadratic fermion couplings, we need only compute to 
order $\epsilon^2$ 

The supersymmetry transformations of the fields of type IIA supergravity 
are listed in appendix B. Acting on the bosonic D0-brane solution 
(\ref{eq:bosonic}), 
we find that the first supersymmetry variations of the 
spinor fields are given by
\bea
\delta \lambda &=& {3 \sqrt{2} \over 8} H^{-{17 \over 16}} \partial_i H 
\Gamma^{i0} P_+ \epsilon \nonumber\\
\delta \psi_0 &=&  - {7 \over 16} H^{-{3 \over 2}} \partial_i H 
\Gamma^{i0} P_+ \epsilon + \left\{ \partial_0 \epsilon \right\} 
\label{eq:fermvar}\\
\delta \psi_i &=&  {1 \over 16} H^{-1} \partial_j H(\Gamma^{ij} - 7 
\delta^{ij}) P_+ \epsilon + \left\{\partial_i \epsilon + {7 \over 32} 
H^{-1} \partial_i H \epsilon \right\} \nonumber
\eea
where $P_+ = {1 \over 2} (1 + \Gamma^0 \Gamma^{11})$ is a projection 
operator. By choosing $\epsilon = H^{-7 \over 32} \eta$ where $\eta$ is 
a constant spinor satisfying $P_+ \eta = 0$, the right hand side of 
these equations vanish, showing that the solution preserves 16 
supersymmetries. We are interested in acting with the broken 
supersymmetry 
generators, which we take to be $\epsilon = H^{-{7 \over 32}} \eta$ with   
$P_+ \eta = \eta$ so that the bracketed terms above vanish.

 Using these fermion transformation rules we may now compute the bosonic 
fields of the superpartner solution to quadratic order in $\epsilon$. We 
find that the long range fields are given by\footnote{In calculating 
these expressions, we take the RR one-form field of the bosonic solution 
to be $C_0 = H^{-1} - 1$. The extra factor of -1 relative to 
(\ref{eq:bosonic}) is physically unimportant since $C$ is a potential, 
but is chosen so that $C$ vanishes at infinity.}
\be
\label{eq:sgfields}
\ba{rllll}
(e_0 {}^i)_{\epsilon^2} & = & {1 \over 2} (\epsilon Q)^2 e_0 {}^i &=& 
-{7i \over 32} H^{-{31 \over 16}} 
\partial_j H \bar{\eta} \Gamma^{ij0} \eta \\
(e_i {}^0)_{\epsilon^2} & = & {1 \over 2} (\epsilon Q)^2 e_i {}^0 &=& {i 
\over 32} H^{-{23 \over 16}} 
\partial_j H \bar{\eta} \Gamma^{ij0} \eta \\
(C_i)_{\epsilon^2} & = & {1 \over 2} (\epsilon Q)^2 C_i &=& {i \over 4} 
H^{-2} \partial_j 
H \bar{\eta} \Gamma^{ij0} \eta \\
(B_{ij})_{\epsilon^2} & = & {1 \over 2} (\epsilon Q)^2 B_{ij} &=& -{i 
\over 4} H^{-1} \partial_k H 
\bar{\eta} \Gamma^{ijk0} \eta \\
(C_{0ij})_{\epsilon^2} & = & {1 \over 2} (\epsilon Q)^2 C_{0ij} &=& {i 
\over 4} H^{-2} \partial_k H 
\bar{\eta} \Gamma^{ijk0} \eta 
\ea
\ee
Note that these are precisely the fields that we found coupling to 
worldvolume fermion bilinears. 

We would like to understand the relationship between the supersymmetry 
variation parameter $\eta$ in these expressions and the worldvolume 
fermion operator $\Theta$ appearing in (\ref{eq:suglin}). From equation 
(\ref{eq:fermvar}), we see that $\eta$ is related to the ``values'' of 
the bulk fermion fields in the superpartner solutions. Quantum 
mechanically, these fermion zero-modes are operator valued, and from the 
supergravity point of view, it is these fermion zero-mode operators that 
are responsible for creating the 256-dimensional multiplet of D0-brane 
states. Thus, the parameters $\eta$ appearing in the superpartner 
solutions should be viewed as operators acting on the Fock space of 
D0-brane states, and (as may be deduced from the bulk fermion 
anticommutation relations) these operators satisfy a Clifford algebra 
equivalent to that of the worldvolume fermions $\Theta$. 

In order to directly relate the parameters $\eta$ with the worldvolume 
fermions considered in 
the previous sections, we need to make a change of variables, since we 
had defined ${1 \over 2} (1 - \Gamma^{11}) \Theta = \Theta$ while the 
broken supersymmetry generators satisfy ${1 \over 2}(1 + \Gamma^0 
\Gamma^{11})\eta = \eta$. The transformation relating these two is 
\[
\eta = {1 \over 4}( \Gamma^{11} + \Gamma^{11} \Gamma^0) \Theta 
\; .
\]
Since the normalization of $\eta$ was arbitrary, we were free to choose 
the overall normalization on the right hand side of this equation. Using 
this transformation, we find
\[
\bar{\eta} \Gamma^{ij0} \eta = {1 \over 2} \bar{\Theta} \Gamma^{0ij} 
\Theta, \qquad 
\qquad \qquad \bar{\eta} \Gamma^{ijk0} \eta = {1 \over 2} \bar{\Theta} 
\Gamma^{ijk} 
\Theta
\]
With this substitution, the long range fields from (\ref{eq:sgfields}) 
precisely match those calculated earlier (\ref{eq:suglin}) from the 
linear couplings.

\section{Classification of D0-brane polarizations}

In the previous sections, we have shown that the action for a single 
D0-brane contains many couplings between the worldvolume fermions and 
the background type IIA supergravity fields. These couplings will cause 
the various polarization states to behave differently in the presence of 
non-zero background fields, and as discussed in the previous section 
lead to different long range supergravity fields. All of the operators 
coupling linearly to bosonic supergravity fields are built out of the 
basic operators
\[
J^{ij} \equiv -{i \over 4} \theta \gamma^{ij} \theta \qquad \qquad 
D^{ijk} \equiv {i \over 4} \theta \gamma^{ijk} \theta
\]
These are also the objects that appear in the operator valued 
expressions for the long range fields of the supergravity solution. 
Thus, all information necessary to understand the physical properties of 
the 
various polarization states is determined by the action of the $J$ and 
$D$ operators on the 256 polarization states. 

In this section, we provide an explicit representation of the 
polarization states. We show that the operators $J$ and $D$ together 
generate SO(16) and that the bosonic and fermionic states lie in 
opposite chirality {\bf 128} spinor representations of this group. 
Finally, we write down 
explicitly the actions of $J$ and $D$ on an arbitrary polarization state 
and use these to classify the states according to their eigenvalues for 
a 
physically interesting maximally commuting set of generators. 

\subsection{Fock space of D0-brane polarizations}

In order to label the D0-brane states, it is useful to rearrange the
$\theta$'s into creation and annihilation operators \bea
\lambda_\alpha &=& {1 \over \sqrt{2}}(\theta_\alpha - i \theta_{8 +
\alpha})
\label{eq:cran}\\
\lambda^\dagger_\alpha &=& {1 \over \sqrt{2}}(\theta_\alpha + i 
\theta_{8 + \alpha}) 
\nonumber
\eea
with $\alpha = 1, \dots, 8$. These obey
\[
\{ \lambda_\alpha ,  \lambda^\dagger_\beta \} = \delta_{\alpha \beta}, 
\; \; \; \; \; 
\{\lambda_\alpha , \lambda _\beta \} = \{ \lambda^\dagger_\alpha, 
\lambda^\dagger_\beta \} 
= 0 .
\]
The polarization states of the D0-brane may then be constructed by
acting with creation operators $\lambda^\dagger$ on a state $|-\rangle$
which is annihilated by all $\lambda$'s. A given state may be labeled
by
\[
| c_1 \cdots c_8 \rangle = (\lambda^\dagger_1)^{ c_1} \cdots 
(\lambda^\dagger_8)^{c_8}|-\rangle
\]
where $c_\alpha \in {0,1}$.
\vskip 0.1 in \noindent 
These $256$ states of a D0-brane may be
understood naturally in terms of their M-theory origins.  They are
simply the polarization states of the 11-dimensional
supergraviton. These form a representation of $SO(9)$, the little
group for massless particles in 11 dimensions\footnote{Of course, this
is also the little group for massive particles in 10 dimensions.}, and
may be divided into three irreducible representations corresponding to
the graviton, three-form, and gravitino of $D=11$ supergravity.\\ 

Let us focus first on the bosonic states. There are a total of $128$ 
independent bosonic states of a D0-brane, $44$ arising from the 
polarization states of the 11-dimensional graviton and $84$ arising from 
the polarization states of the 11-dimensional three-form field. A 
general bosonic state will be some linear combination of these $128$ 
states, and we may represent it by polarization tensors
\[
\{ h_{ij}, \; A_{ijk} \} \qquad \qquad \{i,j,k = 1, \dots, 9 \}
\]
where $h_{ij}$ is a complex symmetric traceless tensor and $A_{ijk}$ is 
a complex antisymmetric tensor. 

The remaining 128 independent D0-brane states are fermionic and arise 
from the 128 polarization states of the $D=11$ gravitino field. The most 
general fermionic D0-brane state is a linear combination of these 
states, and we may represent it by a gravitino polarization tensor
\[
\psi_{i \alpha} \qquad \qquad \{i=1, \dots, 9 ; \; \alpha = 1 \dots 16 
\} 
\]
where $\psi$ is a complex vector-spinor satisfying the 16 constraints 
$\gamma^i_{\alpha \beta} \psi_{i \beta} = 0$. 

An explicit construction of the general states $| h_{ij} , \; A_{ijk} 
\rangle$ and $| \psi_{i \alpha} \rangle$  in terms of the creation and 
annihilation operators $\lambda$ may be found in \cite{pw}. It is 
possible to 
choose conventions such that the inner product between arbitrary states 
is given by 
\be
\label{eq:norm}
\langle \tilde{h}, \; \tilde{A} | h, \; A \rangle = \tilde{h}^*_{ij} 
h_{ij} + \tilde{A}^*_{ijk} A_{ijk} \qquad \qquad \langle \tilde{\psi} | 
\psi \rangle = \tilde{\psi}^*_{i \alpha} \psi_{i \alpha} 
\ee
The two sets of bosonic states constitute the irreducible {\bf 44} and 
{\bf 84} representations of SO(9), the little group for massive 
particles in 10 dimensions, while the fermionic states form a single 
irreducible {\bf 128} representation. 
In this sense, we have three different types of D-particles. However, in 
the presence of non-zero background supergravity fields, the two types 
of bosonic states mix with each other. 

{}From the action (\ref{eq:couplings}), we see that the effects of 
background fields on the D0-brane states, as well as the background 
fields generated by a given state will be governed by the two 
operators
\bea
J^{ij} &\equiv& -{i \over 4} \theta \gamma^{ij} \theta \nonumber\\
D^{ijk} &\equiv& {i \over 4} \theta \gamma^{ijk} \theta \label{eq:JDdef} 
.
\eea
The commutation relations of these operators may be determined in a 
straightforward way from the anticommutation relations for the 
$\theta$'s The operators $J^{ij}$ have the algebra of 
$SO(9)$,\footnote{Here and in the rest of this paper, symmetrization 
(denoted by $(i_1 \cdots i_n))$ and antisymmetrization (denoted by $[i_1 
\cdots i_n])$ of indices are taken with weight 1. For example, $M_{[ij]} 
\equiv {1 \over 2}(M_{ij} - M_{ji})$.} 
\[
[J^{ij}, J_{kl}] = 4i\delta^{[i}{}_{[k} J^{j]}{}_{l]}
\]
This is to be expected since $J^{ij}$ are precisely the fermionic parts 
of the operators which generate spatial rotations in the theory. The 
commutation relations between $D$ and $J$ operators reflect the 
property that the $D$'s are in the 3-index antisymmetric tensor 
representation of the $SO(9)$ generated by $J$'s. They are
\[
[J^{ij}, D_{klm}] = 6i\delta^{[i}{}_{[k} D^{j]}{}_{lm]}
\]
Finally, the commutation relations for the $D$'s are given by
\[
[D^{ijk}, D_{lmn}] = 18i\delta^{[i}{}_{[l} \delta^j{}_m J^{k]} {}_{n]} - 
{i \over 6} 
\epsilon^{ijklmnpqr} D_{pqr}
\]
It turns out that the $J$'s and $D$ together generate $SO(16)$, however, 
they form a somewhat unusual set of generators. The standard $SO(16)$ 
generators are simply the fermion bilinears
\[
A_{\alpha \beta} = \theta_\alpha \theta_\beta 
\]
as may be easily checked. The $J$'s and $D$'s form a different basis of 
the $120$ independent bilinears, and we see from (\ref{eq:JDdef}) that 
the matrices $\gamma^{ij}_{\alpha \beta}$ and $\gamma^{ijk}_{\alpha 
\beta}$ are the coefficients which relate the ordinary basis to the 
$J,D$ basis. 

The bosonic D0-brane states which formed a {\bf 44} and {\bf
84} of $SO(9)$ combine into a single {\bf 128} chiral spinor 
representation of this $SO(16)$.  It is interesting to note that this 
representation of $SO(16)$ is precisely the one which appears when we 
consider the action of $SO(16)$ on the coset space $E_8/SO(16)$
\cite{Adams-exceptional}.  This suggests that perhaps the D0-brane
polarization states may correspond to broken symmetry generators in
some more symmetric phase of string theory, but we will not pursue
this connection further here. The fermionic states also lie in a {\bf 
128} chiral spinor representation of $SO(16)$, but of the opposite 
chirality. 

\subsection{Action of spin and dipole operators on polarization states}

The action of the $SO(16)$ generators $J$ and $D$ on the 
general bosonic and fermionic states may be determined using the 
explicit representation of the states found in \cite{pw}. We first 
determine 
the action of a single operator $\theta_\alpha$ on the general state. We 
find
\bea
\theta_\alpha: |h, \; A \rangle &\rightarrow& |\tilde{\psi} \rangle 
\nonumber \\
\tilde{\psi}_{i \beta} &=& { \sqrt{2} \over 2} \gamma^j_{\beta \alpha} 
h_{ij} - {\sqrt{6} \over 36} (\gamma^{ijkl}_{\beta \alpha} - 6 
\delta^{ij} \gamma^{kl}_{\beta \alpha})A_{jkl} \label{eq:Thact} \\
\theta_\alpha: |\psi \rangle &\rightarrow& |\tilde{h}, \; \tilde{A} 
\rangle \nonumber \\
\tilde{h}_{ij} &=& {\sqrt{2} \over 2} \gamma^{(i}_{\alpha \beta} 
\psi^{j)}_\beta \nonumber\\
\tilde{A}_{ijk} &=& -{\sqrt{6} \over 4} \gamma^{[ij}_{\alpha \beta} 
\psi^{k]}_\beta \nonumber
\eea
Using these relations, we find (as expected) that the rotation 
generators act as
\bea
J^{ij}: |h, \; A \rangle &\rightarrow& |\tilde{h}, \; 
\tilde{A} \rangle \nonumber \\
\tilde{h}_{kl} &=& -4i \delta^{[i}{}_{(k} h^{j]} {}_{l)} 
\label{eq:Jact}\\
\tilde{A}_{klm} &=& -6i \delta^{[i} {}_{[k} A^{j]} {}_{lm]} \; . 
\nonumber \\
J^{ij}: |\psi \rangle &\rightarrow& |\tilde{\psi} \rangle \nonumber \\
\tilde{\psi^k} &=& -2i \delta_k {}^{[i} \psi^{j]} - {i \over 2} 
\gamma^{ij} \psi^{k} \nonumber
\eea
The operators $D$ mix states arising from graviton and three-form 
polarizations. We have
\bea
D^{ijk}: |h, \; A \rangle &\rightarrow& |\tilde{h}, \; 
\tilde{A} \rangle \nonumber\\
\tilde{h}_{lm} &=& 6i \sqrt{3} \left( \delta^{[i}{}_{(l}  A^{jk]} 
{}_{m)} - {1 \over 9} \delta_{lm} A^{ijk} \right) \label{eq:Dact}\\
\tilde{A}_{lmn} &=& -6i \sqrt{3} \delta^{[i} {}_{[l} \delta^j {}_m  
h^{k]} {}_{n]} - {i \over 6} \epsilon^{ijklmnpqr} A_{pqr} \; . \nonumber
\\ 
D^{ijk}: |\psi \rangle &\rightarrow& |\tilde{\psi} \rangle \nonumber \\
\tilde{\psi^l} &=& -4i \delta_l {}^{[i} \gamma^j \psi^{k]} + i 
\gamma^{l[ij} \psi^{k]} - {i \over 2} \gamma^{ijk} \psi^{l} \nonumber
\eea
The relations (\ref{eq:Jact}) and (\ref{eq:Dact}), together with the 
inner product (\ref{eq:norm}) may be used to compute the action of an 
arbitrary operator built from $\theta$s on a general bosonic state, as 
well as the expectation values of arbitrary operators.

For example, we find
\beas
\langle J^{ij} \rangle &=& 4 \; {\rm Im}( h^*_{il} h_{jl} ) + 6 \; {\rm 
Im} (A^*_{ilm} A_{jlm})\\
\langle D^{ijk} \rangle &=& 12 \sqrt{3} \;{\rm Im}( h_{m[i} A^*_{jk]m} ) 
+ {1 \over 3} \; {\rm Im} (\epsilon^{ijklmnpqr} A^*_{lmn} A_{pqr})\\
\eeas
Recalling the couplings (\ref{eq:couplings}), we see that the first of 
these expressions 
gives the expectation value of the angular momentum in the $\{ij\}$ 
plane (equal to the RR one-form magnetic moment), while the second gives 
the D2-brane dipole moment in the $\{ijk\}$ directions (equal to the 
NS-NS two form magnetic moment). These expectation values are exactly 
those needed to evaluate the long range dipole fields (\ref{eq:suglin}) 
for a given polarization state.

\subsection{Classification in terms of $J$ and $D$ eigenstates.}

To understand the physical properties of the various D0-brane 
states, it is useful to choose a maximal mutually commuting set of 
generators (Cartan subalgebra) and then write the states in a basis of 
simultaneous eigenstates for these generators. 

A physically interesting choice of commuting generators is the set 
\[
\{ J_1 \equiv J^{12}, J_3 \equiv J^{34}, J_5 \equiv J^{56}, J_7 \equiv 
J^{78}, D_1 \equiv D^{129}, D_3 \equiv D^{349}, D_5 \equiv D^{569}, D_7 
\equiv D^{789} \}
\]
containing four rotation operators and four dipole operators.

We may now describe an orthogonal basis of the bosonic and fermionic 
states which 
are simultaneous eigenstates of these 8 generators. A useful property of 
these basis elements is that the expectation values of all other 
generators not in the Cartan subalgebra vanish\footnote{This follows 
since all other generators may be written as the commutator of some 
generator with a Cartan subalgebra generator.}. 

\subsubsection{Basis for bosonic states}

We begin with the 128 independent bosonic states. We will write bosonic 
eigenstates in terms of a simpler (non-normalized and non-orthogonal) 
basis given by
\begin{eqnarray*}
|ij\rangle & \equiv & |h_{ij} = h_{ji}= 1\rangle,\;\;\;\;\;i \ne j \\
|ii\rangle & \equiv & |h_{ii} = -h_{99} = 1 \rangle, \;\;\;\;\;i=1, 
\dots ,8\\
|ijk\rangle &\equiv& |A_{ijk} = -A_{ikj} = A_{jki} = -A_{jik} = A_{kij} 
= -A_{kji} = 1\rangle
\end{eqnarray*}

The action of each of the 8 Cartan generators on this basis may be read 
off from the relations (\ref{eq:Jact}) and (\ref{eq:Dact}). Using these, 
we first diagonalize the $J$'s and then diagonalize the $D$'s in each 
subspace of states of fixed $J$'s. Below we will use the indices 
$a,b,c,d$ to represent distinct elements of $\{1,3,5,7\}$, labeling the 
generators in our Cartan subalgebra. Also, for $a=2l-1$, we let 
$\hat{a}=2l$. 
\vskip 0.1 in  \noindent
{\bf group 1: $J =0$ $D = \pm2$}
\vskip 0.1 in  \noindent  
There are 8 basis states for which all $J$'s vanish. These include four 
graviton states 
\[
|a_0\rangle  \equiv |aa\rangle + |\hat{a} \hat{a}\rangle
\]
and four three-form states
\[
|a 9\rangle \equiv |a \hat{a} 9\rangle.
\]
Diagonalizing the $D$ generators on the subspace generated by these 
states, we find that the diagonal basis consists of states for which a 
single $D$ generator has the eigenvalue $\pm 2$. Explicitly, the 
normalized eigenstates are:
\[
| D_a = \pm2 \rangle \equiv {1 \over 6}\left( 2|a_0 \rangle - |b_0 
\rangle - |c_0 \rangle - |d_0 \rangle \mp i\sqrt{3} |a 9 \rangle \right)
\]
\vskip 0.1 in  \noindent
{\bf group 2: $J = \pm 2$, $D=0$}
\vskip 0.1 in  \noindent
There are $8$ states for which a single $J_a$ has a value of $\pm2$. 
These all come from 
graviton polarizations and are given by
\[
|J_a = {\pm 2}\rangle = {1 \over 2} |a \hat{a} \rangle \mp {i \over 2} 
|aa\rangle \pm {i 
\over 2} |\hat{a} \hat{a}\rangle.
\]
For these states, $J_a = \pm2$ and all other Cartan generators including 
$D$s vanish
\vskip 0.1 in  \noindent
{\bf group 3: $J_a = \pm 1$,  $D_b, D_c, D_d = \pm 1$}
\vskip 0.1 in  \noindent
There are 32 states for which a single $J_a$ has the value $\pm 1$. 
These include 8 graviton states
\[
|a_{\pm 1} 9\rangle \equiv |a 9\rangle \pm i |\hat{a} 9\rangle
\]
as well as 24 three-form states
\[
|a_{\pm 1} b\rangle \equiv |a b \hat{b}\rangle  \pm i |\hat{a} b 
\hat{b}\rangle
\]
These are not eigenstates of the $D_a$'s, however we may combine them 
into normalized eigenstates
\[
|J_a = \epsilon, \; D_b = \lambda_b, \; D_c= \lambda_c,  \; D_d = 
\lambda_d \rangle  \equiv  {1 \over 4} 
|a_{\epsilon} 9 \rangle - {i \lambda_b \over 4 \sqrt{3} } |a_{\epsilon} 
b\rangle - {i \lambda_c \over 4 \sqrt{3} } |a_{\epsilon} c\rangle  - {i 
\lambda_d \over 4 \sqrt{3} } 
|a_{\epsilon} d\rangle
\]
where $\epsilon, \lambda_b \lambda_c, \lambda_d \in {\pm 1}$ are the 
eigenvalues of 
$J_a, D_b, D_c $, and  $D_d$ respectively, with the constraint that 
$\lambda_b \lambda_c 
\lambda_d = \epsilon$. Recalling that $a,b,c,d$ must be distinct, we may 
verify that this gives a total of 32 states. 
\vskip 0.1 in  \noindent
{\bf group 4: $J_a = \pm 1, J_b = \pm 1$, $D_a, D_b = \pm 1$}
\vskip 0.1 in  \noindent
There are 48 states for which two different $J_a$'s have (uncorrelated) 
eigenvalues of $\pm 1$. These include 24 graviton states
\[
|a_{\epsilon_a} b_{\epsilon_b}\rangle \equiv |a b \rangle + i 
\epsilon_a |\hat{a} b\rangle + i \epsilon_b | a \hat{b}\rangle - 
\epsilon_a \epsilon_b |\hat{a} \hat{b}\rangle
\]
and 24 three-form states
\[
|a_{\epsilon_a} b_{\epsilon_b} 9\rangle \equiv |a b 9\rangle  + i 
\epsilon_a |\hat{a} b 9\rangle + i \epsilon_b | a \hat{b} 9\rangle - 
\epsilon_a \epsilon_b |\hat{a} \hat{b} 9\rangle
\]
where $\epsilon_a,\epsilon_b \in {\pm1}$ are the eigenvalues for $J_a$ 
and $J_b$. These states are mixed by the $D_a$'s but we may 
combine them into eigenstates as
\[
|J_a = {\epsilon_a}, \; J_b = {\epsilon_b}, \; D_a = \epsilon_a \delta, 
\; D_b = -\epsilon_b \delta \rangle \equiv {1 \over 4} 
|a_{\epsilon_a} b_{\epsilon_b}\rangle 
+ {\delta  \over 4 \sqrt{3}} |a_{\epsilon_a} b_{\epsilon_b} 9 \rangle
\]
where $\delta = \pm 1$. 
\vskip 0.1 in  \noindent
{\bf group 5: $J_a = \pm 1, J_b = \pm 1, J_c = \pm 1, D_d = \pm 1$}
\vskip 0.1 in  \noindent
The remaining 32 states have three distinct $J_a$'s equal to $\pm 1$. 
These states all arise from three-form polarizations and are given by
\[
|J_a = {\epsilon_a}, \; J_b = {\epsilon_b}, \; J_c = {\epsilon_c}, \;  
D_d= \epsilon_a \epsilon_b \epsilon_c \rangle = {1 \over 4 \sqrt 
3}\left(|a b c \rangle + i \epsilon_a |\hat{a} b c\rangle + \dots - i 
\epsilon_a \epsilon_b \epsilon_c | \hat{a} \hat{b} \hat{c}\rangle 
\right)
\]
Note that each of these states is also an eigenstate of the $D$s with 
$D_a = D_b = D_c = 0$ and $D_d = \epsilon_a \epsilon_b \epsilon_c$

\subsubsection{Fermionic states}

To describe the fermionic eigenstate basis, we introduce projection 
operators
\[
P^a_\pm = {1 \over 2}(1 \pm i \gamma^{a \hat{a}})
\]
where, as above, $a \in \{1,3,5,7\}$ and $\hat{a} = a+1$. Each of these 
independently reduces the number of independent components of a
16-component spinor by half, so the state $P^1_{\epsilon_1}  
P^3_{\epsilon_3} P^5_{\epsilon_5} P^7_{\epsilon_7} \chi$ has only a 
single independent component. Each state in the fermionic basis has an 
eigenvalue of $\pm{3 \over 2}$ for exactly one of the 8 Cartan 
subalgebra generators, with eigenvalues of $\pm{1 \over 2}$ for the 
remaining 7 generators. We now describe these states explicitly.
\\ \\
{\bf group 1:} 
\\ \\
The first set of 64 states has eigenvalues
\beas
J_a = {3 \over 2} \epsilon_a \qquad J_b = {1 \over 2} \epsilon_b 
&\qquad& J_c = {1 \over 2} \epsilon_c \qquad J_d = {1 \over 2} 
\epsilon_d \\
D_a = -{1 \over 2} \epsilon_b \epsilon_c \epsilon_d \qquad D_b = -{1 
\over 2} \epsilon_a \epsilon_c \epsilon_d &\qquad& D_c = -{1 \over 2} 
\epsilon_a \epsilon_b \epsilon_d \qquad D_d = -{1 \over 2} \epsilon_a 
\epsilon_b \epsilon_c 
\eeas
where as above, $\{a,b,c,d\} = \{1,3,5,7\}$. Since each $\epsilon$ can 
be $\pm 1$ and we have 4 choices for $a$, this gives 64 states, so the 
states are specified uniquely by their eigenvalues. The polarization 
vectors for these basis elements are given by
\[
\psi^a = -i \epsilon_a \psi^{\hat{a}} = P^a_{-\epsilon_a} 
P^b_{-\epsilon_b} P^c_{-\epsilon_c} P^d_{-\epsilon_d} \chi, \qquad 
\psi^i = 0 \qquad \{i \ne a, \hat{a} \}
\]
Note that the choice of $\chi$ is irrelevant since there is only a 
single independent component after acting with the four projection 
operators.
\\ \\
{\bf group 2:}
\\ \\
The remaining 64 fermionic states have eigenvalues
\beas
J_a = {1 \over 2} \epsilon_a \qquad J_b = {1 \over 2} \epsilon_b 
&\qquad& J_c = {1 \over 2} \epsilon_c \qquad J_d = {1 \over 2} 
\epsilon_d \\
D_a = {3 \over 2} \epsilon_b \epsilon_c \epsilon_d \qquad D_b = -{1 
\over 2} \epsilon_a \epsilon_c \epsilon_d &\qquad& D_c = -{1 \over 2} 
\epsilon_a \epsilon_b \epsilon_d \qquad D_d = -{1 \over 2} \epsilon_a 
\epsilon_b \epsilon_c 
\eeas
Again, these states are specified uniquely by their eigenvalues. The 
polarization vectors for these basis elements are given by 
\beas
\psi^a &=& -i \epsilon_a \psi^{\hat{a}} = P^a_{+\epsilon_a} 
P^b_{-\epsilon_b} P^c_{-\epsilon_c} P^d_{-\epsilon_d} \chi\\
\psi^b &=& -i \epsilon_b \psi^{\hat{b}} = -{1 \over 2} P^a_{-\epsilon_a} 
P^b_{+\epsilon_b} P^c_{-\epsilon_c} P^d_{-\epsilon_d} \gamma^{ba} \chi\\
\psi^c &=& -i \epsilon_c \psi^{\hat{c}} = -{1 \over 2} P^a_{-\epsilon_a} 
P^b_{-\epsilon_b} P^c_{+\epsilon_c} P^d_{-\epsilon_d} \gamma^{ca} \chi\\
\psi^d &=& -i \epsilon_d \psi^{\hat{d}} = -{1 \over 2} P^a_{-\epsilon_a} 
P^b_{-\epsilon_b} P^c_{-\epsilon_c} P^d_{+\epsilon_d} \gamma^{da} \chi\\
\psi^9 &=& \epsilon_a \epsilon_b \epsilon_c \epsilon_d \gamma^a \psi^a
\eeas

\section{Static interactions between polarized D0-branes}

In this section, we apply our results and consider the long-range 
interactions between a
pair of D0-branes in fixed polarization states.  In particular, we
would like to consider the interactions between a pair of D0-branes in
identical polarization states.  Since each D0-brane is in a state
which preserves the same 16 supersymmetries, a pair of D0-branes in
identical polarization states should be a BPS configuration in which
the force between the branes is identically 0.

The long-range force between a pair of D0-branes with spin and
magnetic D0 dipole moment was considered in \cite{dlr}.  These authors
showed that the spin-spin and dipole-dipole interactions precisely
cancel since the gyromagnetic ratio of the D0-brane is $g = 1$.  We
show here that this result follows immediately from the couplings
(\ref{eq:couplings}), and furthermore that the dipole-dipole
interactions mediated by the R-R 3-form and NS-NS 2-form field also
cancel identically due to the equality of the D2-brane dipole and
magnetic H-dipole moments.

The cancellation of all dipole-dipole forces between static D0-branes 
follows immediately from the formulae for long range fields derived in 
the previous section. We simply let one of the D-branes act as a source 
which generates the long range fields calculated in the previous 
section, and treat the other brane as a probe which feels a potential 
obtained by plugging in the fields generated by the first brane into the 
action (\ref{eq:couplings2}). Examining the expressions 
(\ref{eq:suglin}) and (\ref{eq:couplings2}), it is obvious that 
the spin-spin potential mediated by the graviton
\[
V_{\rm spin} = {15 \over 32} \pi^3 i \bar{\Theta}_1 \Gamma^{0ij} 
\Theta_1 \bar{\Theta}_2 \Gamma^{0ik} \Theta_2  \partial_j \partial_k 
\left\{ {1 \over r^7} \right\}
\]
is precisely canceled by the RR one-form mediated dipole-dipole 
potential
\[
V_{\rm dipole} = -{15 \over 32} \pi^3 i \bar{\Theta}_1 \Gamma^{0ij} 
\Theta_1 \bar{\Theta}_2 \Gamma^{0ik} \Theta_2  \partial_j \partial_k 
\left\{ {1 \over r^7} \right\} \; ,
\]
as was demonstrated previously in \cite{dlr}. In a similar way, the 
interaction between D2-brane dipole moments gives rise to a potential
\[
V_{C^{(3)}} = {15 \over 32} \pi^3 \bar{\Theta}_1 \Gamma^{ijk} \Theta_1 
\bar{\Theta}_2 \Gamma^{ijl} \Theta_2 \partial_k \partial_l \left\{ {1 
\over r^7} \right\}
\]
which is precisely canceled by the NS-NS two-form mediated force 
between magnetic H-dipole moments,
\[
V_{B} = -{15 \over 32} \pi^3 \bar{\Theta}_1 \Gamma^{ijk} \Theta_1 
\bar{\Theta}_2 \Gamma^{ijl} \Theta_2 \partial_k \partial_l \left\{ {1 
\over r^7} \right\}
\]
Recalling that the leading order gravitational and dilaton mediated 
forces 
also cancel precisely with the repulsive force between the two D0-brane 
charges, we see that the forces between two static D0-branes cancel 
completely up to order ${1 \over r^9}$. 

It is interesting that we did not need to make any assumptions about the 
polarizations of the two D0-branes. We should point out, however, that 
we do not expect all forces between two static D0-branes to cancel in 
general. In fact, the general expression for the long range static 
potential has been worked out \cite{bhp,psw}, and is given by
\[
V_{D0-D0} = -{5 \over 43008} \theta \gamma^{mi} \theta \theta 
\gamma^{mj} 
\theta \theta 
\gamma^{nk} \theta \theta \gamma^{nl} \theta \; \partial_i \partial_j 
\partial_k \partial_k \left\{ {1 \over r^7} \right\}
\]
where $\theta = \theta_1 - \theta_2$ is the relative polarization of the 
two D0-branes in the 16-component spinor notation. This is a ${1 \over 
r^{11}}$ potential, which should be reproducible using our results 
through a 
combination of quadrupole-quadrupole, octupole-dipole and 
16pole-monopole forces. It is only when the two D0-branes have the same 
polarization that we expect to obtain a classical BPS state and thus 
observe the complete cancellation of all forces. 

\section{Discussion}

In this paper, we have investigated the physical properties of the 
different polarization states of a D0-brane. By analyzing the couplings 
of the D0-brane worldvolume fermions to the type IIA supergravity bulk 
fields, we have seen that in addition to mass and RR one-form charge, 
D0-branes may carry moments of a variety of other conserved quantities. 
The 
dipole moments include angular momentum (spin), RR one-form magnetic 
moment, D2-brane dipole moment, and NS-NS two-form magnetic dipole 
moment. We have shown a complete cancellation between dipole-dipole 
forces for a pair of static D0-branes owing to the fact that D0-branes 
have a gyromagnetic ratio of 1 and that the ratio between D2-brane 
dipole moment and NS-NS two-form magnetic moment is also 1. 

We have determined the leading long range supergravity 
fields corresponding to an arbitrary polarization state of the D0-brane. 
For each of these states, there should also be a corresponding exact 
supergravity solution, thus we expect there to be a many parameter 
family of black hole solutions in $D=10$ preserving 16 supersymmetries 
and carrying multipole moments for various types of conserved 
quantities. For example, we expect supergravity solutions corresponding 
to each element of the eigenstate basis of section 4, including 
solutions
with zero (classical) angular momentum but with dipole moments of 
D2-brane charge. It would be interesting to find these exact solutions 
using the knowledge of their long range fields.

One approach to deriving the exact supergravity solutions would be to 
work in the context of $D=11$ supergravity. Since type IIA supergravity 
is related to eleven-dimensional supergravity by dimensional reduction, 
any exact solution of type IIA supergravity is related to a solution of 
$D=11$ supergravity with translational invariance along one of the 
spatial directions. In this way, the bosonic D0-brane solution of type 
IIA supergravity is related to the Aichelburg-Sexl solution of $D=11$ 
supergravity \cite{Aichelburg-Sexl} corresponding to the gravitational 
fields around a massless 
particle.\footnote{More precisely, the lifted D0-brane solution becomes, 
after a coordinate transformation, the Aichelburg-Sexl solution smeared 
along the direction of particle motion.} The superpartners of the 
D0-brane solution should then be identified with superpartners of the 
Aichelburg-Sexl solution, and these should take a particularly simple 
form since the Aichelburg-Sexl solution has only a single non-vanishing 
field, $h_{--} \propto {1 \over r^7}$. Given the Aichelburg-Sexl 
superpartner solutions, the complete D0-brane superpartner solutions 
would then be obtained by reducing to $D=10$ in the standard way.

The couplings of bulk fields to worldvolume fermions that we have 
derived have interesting implications for the study of systems of two 
types of branes, for example in the investigation of bound states 
between D0-branes and other types of branes. If we consider one set of 
branes to be a source for various bulk fields, the couplings of these 
fields to the fermions on the worldvolume of the probe D0-branes will 
result in different energies for the different polarization states of 
the D0-branes. The minimum energy configuration will likely be one in 
which the D0-branes lie in a specific polarization state, and the long 
range supergravity fields corresponding to this system of branes should 
exhibit any non-zero moments associated with this polarization state.
This may be related to an interesting observation made in 
\cite{Sethi-Stern04}. 
In that paper, it was found that the wavefunction of a D0-brane in the 
D0-D4 bound state is indicative of a specific combination of 
polarization states 
(representations of $Spin(5)$ in their case).\footnote{We thank Savdeep 
Sethi 
for pointing this out to us.} It is possible that an understanding of 
this 
phenomenon would result from considering the couplings of D0-brane 
fermions to 
the supergravity fields generated by the D4-brane.
 
Finally, we note that almost identical considerations apply to the
various other types of branes in theories with 32 supersymmetries (as
well as theories with less supersymmetry). Just like the type IIA
D0-brane, configurations of parallel BPS branes in these theories
generally preserve half the supersymmetry of the relevant theory. The
remaining 16 broken supersymmetry generators generate a BPS multiplet
of 256 states which splits into two {\bf 128} representations of the
$SO(16)$ generated by bilinears of the broken generators. In the case
where the spatial dimensions of the brane are compactified, these
branes may be viewed as particle states from the point of view of the
uncompactified directions, and the various states will appear as
different polarizations of some number of particle types
(representations of the rotation group of the uncompactified
space). For particles obtained from toroidally compactified Dp-branes,
the physical properties of the various polarization states can be
investigated using a similar approach to that used in this paper. The
couplings of 
worldvolume fermions to background fields for these particles may be
obtained from the fermion terms in the Dp-brane actions derived in
\cite{Mark-Wati-5}. These more general particle states will correspond
to a large class of black hole solutions in various dimensions, and it
would be interesting to investigate their properties.  One potentially
interesting related direction for further work would be to use the
methods of this paper to study the breaking of degeneracy between
fermionic states in the world-volume theory on a system of 3-branes
forming a dielectric 5-brane sphere (as studied, for example, in
\cite{Polchinski-Strassler}).

\section*{Acknowledgments}

We would like to thank Andrew Chamblin, Marc Grisaru, Marcia Knutt,
Jason Kumar and Savdeep Sethi for discussions.  MVR would like to
thank the physics department at Stanford University for hospitality
during part of this work.  WT would like to thank the Aspen Center for
Physics for hospitality as this paper was being completed.  The work
of MVR is supported in part by NSF grant PHY-9802484. The work of WT
is supported in part by the A.\ P.\ Sloan Foundation and in part by
the DOE through contract \#DE-FC02-94ER40818.  The work of KM is
supported in part by the DOE through contract \#DE-FC02-94ER40818.

\appendix

\section{Properties of spinors}

In this paper we have used two different types of spinors, 16-component 
spinors denoted by $\theta$ and 32-component Majorana-Weyl fermions 
denoted by $\Theta$. With our conventions, the spinors are related by
\[
\Theta = \left( \ba{c} 0 \\ \theta \ea \right) 
\]
For the 16-component spinors, we use a real symmetric set of $16 \times 
16$ Dirac matrices denoted by $\gamma^i$, $ i = 1\dots9$. The $32 \times 
32$ Dirac matrices denoted by $\Gamma$ are given in terms of the 
$\gamma$'s by 
\[
\Gamma^i = \left( \ba{cc} 0 & \gamma^i \\ \gamma^i & 0 \ea \right) 
\qquad \qquad \Gamma^0 = \left( \ba{cc} 0 & -1 \\ 1 & 0 \ea \right)
\qquad \qquad \Gamma^{11} = \left( \ba{cc} 1 & 0 \\ 0 & -1 \ea \right)
\]
Because of the anticommutation relations obeyed by the fermions, the 
only nonvanishing fermion bilinears are
\[
\theta \gamma^{ijk} \theta,  \qquad \theta \gamma^{ij} \theta, 
\qquad \theta \gamma^{ijklmn} \theta = -{1 \over 3!} 
\epsilon^{ijklmnpqr} \theta \gamma^{pqr} \theta,  \qquad \theta 
\gamma^{ijklmnp} \theta = -{1 \over 2} \epsilon^{ijklmnpqr} \theta 
\gamma^{qr} \theta
\]
in the 16-component notation and 
\[
\bar{\Theta} \Gamma^{abc} \Theta \qquad \qquad \bar{\Theta} 
\Gamma^{abcdefg} \Theta = {1 \over 3!} \epsilon^{abcdefghij} 
\bar{\Theta} \Gamma_{hij} \Theta
\]
for 32-component spinors.
In dealing with expressions containing more than two fermions, various 
Fierz identities must be taken into account. For the 16-component 
spinors, a list of these identities may be found in \cite{Mark-Wati-3, 
Hyun-susy}.  For the $D=10$ 
fermions, we have
\[
\bar{\Theta} \Gamma^{abc} \Theta \bar{\Theta} \Gamma_{ab}{}^d \Theta = 0
\]
\[
\bar{\Theta} \Gamma_a{}^{b[c} \Theta \bar{\Theta} \Gamma^{de]a} \Theta  
= 0
\]
\beas
12\bar{\Theta} \Gamma_{abc} \Theta \bar{\Theta} \Gamma^{def} \Theta +
\bar{\Theta} \Gamma_{abcqp}{}^{[ef} \Theta \bar{\Theta} \Gamma^{d]pq} 
\Theta + 
\bar{\Theta} \Gamma^{defqp}{}_{[ab} \Theta \bar{\Theta} \Gamma_{c]pq} 
\Theta & &\\
+ 12 \bar{\Theta} \Gamma_{[ab}{}^{[d} \Theta \bar{\Theta} 
\Gamma_{c]}{}^{ef]} \Theta 
-12 \delta_{[a} {}^{[d} \bar{\Theta} \Gamma_{bc]p} \Theta \bar{\Theta} 
\Gamma^{ef]p} \Theta 
-24 \bar{\Theta} \Gamma_{p[a}{}^{[d} \Theta \delta_{b}{}^{e}\bar{\Theta} 
\Gamma_{c]} {}^{f]p} \Theta && = 0
\eeas
It is important to note that these identities apply for classical 
anticommuting fermions. For operators satisfying anticommutation 
relations $\{ \Theta_\alpha, \Theta_\beta \} = \delta_{\alpha \beta}$, 
we should replace the right hand sides of the first and third 
expressions by $(-320 \; \eta^{cd} - 896  \;\eta^{0c} \eta^{0d})$ and 
$(-768 \; \delta_{[a} {}^{[d} \delta_b {}^e \delta_{c]} {}^{f]} - 4608  
\; \delta^0 {}_{[a} \delta^{[d} {}_b \delta^e {}_{c]} \delta^{f]} {}_0)$ 
respectively. 

\section{Type IIA supergravity conventions}

The fields of $D=10$ type IIA supergravity are the vielbein $e_\mu 
{}^m$, the dilaton $\phi$, the NS-NS two form $B_{\mu \nu}$, the RR one 
and three forms $C_\mu$ and $C_{\mu \nu \lambda}$, the dilatino 
$\lambda$ and the gravitino $\psi_{\mu}$. The supersymmetry 
transformations for these fields are \cite{Huq-Namazie, Campbell-West}
\beas
\delta e_\mu {}^m &=& i \bar{\epsilon} \Gamma^m \psi_\mu\\
\delta \phi &=& i \sqrt{2} \bar{\epsilon} \Gamma^{11} \lambda\\
\delta C_\mu &=& i e^{-{3 \over 4} \phi} \bar{\epsilon} \Gamma^{11} 
\psi_\mu + {3i \sqrt{2} \over 4} e^{-{3 \over 4} \phi} \bar{\epsilon} 
\gamma_\mu \lambda\\
\delta B_{\mu \nu} &=& 2i e^{ \phi \over 2} \bar{\epsilon} 
\gamma_{[\mu} 
\Gamma^{11} \psi_{\nu]} - {\sqrt{2} \over 2}i e^{\phi \over 2} 
\bar{\epsilon} \gamma_{\mu \nu} \lambda \\
\delta C_{\mu \nu \lambda} &=& -3ie^{- {\phi \over 4}} \bar{\epsilon} 
\gamma_{[\mu \nu} \psi_{\lambda]} + 6 i e^{\phi \over 2} C_{[\mu} 
\bar{\epsilon} \gamma_{\nu} \Gamma^{11} \psi_{\lambda]}\\
& &+{\sqrt{2} \over 4}i e^{- {\phi \over 4}} \bar{\epsilon} \gamma_{\mu 
\nu \lambda} \Gamma^{11} \lambda - {3 \sqrt{2} \over 2} i e^{\phi \over 
2}
C_{[\mu}\bar{\epsilon} \gamma_{\nu \lambda]} \lambda\\
\delta \lambda &=& -{ \sqrt{2} \over 4} \gamma^\mu \Gamma^{11} 
 \epsilon \partial_\mu \phi - {3 \sqrt{2} \over 32} e^{{3 \over 4} \phi} 
\gamma^{\mu \nu}  \epsilon F_{\mu \nu} \\
& &+ {\sqrt{2}\over 48}e^{-{1 \over 2} \phi} \gamma^{\mu \nu \lambda} 
\epsilon H_{\mu \nu \lambda} + {\sqrt{2} \over 384} e^{{1 \over 4} \phi} 
\gamma^{\mu \nu \lambda \sigma} \Gamma^{11} F'_{\mu \nu \lambda \sigma} 
+ \dots\\
\delta \psi_\mu &=& \partial_\mu \epsilon + {1 \over 4} \omega_{\mu mn} 
\Gamma^{mn} \epsilon - {1 \over 64} e^{{3 \over 4} \phi} \left\{ 
\gamma_\mu {}^{\nu \lambda} - 14 \delta_\mu {}^\nu \gamma^\lambda 
\right\} \Gamma^{11}  \epsilon F_{\nu \lambda}\\
&& + {1 \over 96} e^{-{1 \over 2} \phi} \left\{ 
\gamma_\mu {}^{\nu \lambda \rho} - 9 \delta_\mu {}^\nu \gamma^{\lambda 
\rho}
\right\} \Gamma^{11}  \epsilon H_{\nu \lambda \rho} + {1 \over 256} 
e^{{1 \over 4} \phi} \left\{ \gamma_\mu {}^{\nu \lambda \rho \sigma} - 
{20 \over 3} \delta_\mu {}^\nu \gamma^{\lambda \rho \sigma} \right\} 
\epsilon F'_{\nu \lambda \rho \sigma} + \cdots
\eeas
Here, Dirac matrices with tangent space indices are denoted by $\Gamma$ 
while Dirac matrices with spacetime indices are denoted by $\gamma$ (not 
to be confused with the $16 \times 16$ Dirac matrices used elsewhere), 
so that
\[
\gamma_\mu = e_\mu {}^m \Gamma_m
\]
Note that the supersymmetry variations of the fermionic fields contain 
additional terms cubic in the fermion fields which are not relevant for 
our purposes. The additional terms may be found in \cite{Huq-Namazie, 
Campbell-West}. Finally, the 
field strengths are defined as
\beas
F_{\mu \nu} &=& 2\partial_{[\mu} C_{\nu]}\\
H_{\mu \nu \lambda} &=& 3 \partial_{[\mu } B_{\nu \lambda]}\\
F'_{\mu \nu \lambda \sigma} &=& 4 \partial_{[\mu} C_{\nu \lambda 
\sigma]} + 8 C_{[\mu} H_{\nu \lambda \sigma]}       
\eeas

\bibliographystyle{plain}


\end{document}